\documentclass{article}

\usepackage{arxiv}

\usepackage[utf8]{inputenc} 
\usepackage[T1]{fontenc}    
\usepackage{hyperref}       
\usepackage{url}            
\usepackage{booktabs}       
\usepackage{amsfonts}       
\usepackage{nicefrac}       
\usepackage{microtype}      
\usepackage{lipsum}
\usepackage{graphicx}
\graphicspath{ {./images/} }

\title{Approximate  Approach to Compute Characteristics of Inhomogeneous TASEP with Open Boundaries
}

\author{
Marina V. Yashina  \\
  Department of Engineering and Mathematics of Applied Artificial Intelligence Systems\\
  Moscow Automobile and Road Construction\\
  State Technical University (MADI) \\
  Moscow, Leningradsky avenue, 64, Russia  \\
  \texttt{mv.yashina@madi.ru} \\ 
\And
 Alexander G. Tatashev  \\
  Department of Engineering and Mathematics of Applied Artificial Intelligence Systems\\
  Moscow Automobile and Road Construction\\
  State Technical University (MADI) \\
  Moscow, Leningradsky avenue, 64, Russia  \\
  \texttt{a-tatashev@yandex.ru} \\
}

\begin{document}
\maketitle

\large

\begin{abstract}
A discrete-time totally asymmetric simple exclusion process on a lattice with open boundaries is considered. There are particles of different types. The type of a particle is characterized by the probability that a particle moves to a vacant site and the probability that a particle occupying the rightmost site departs the system.  An approximate approach to compute the particle flow rate and density in sites is proposed. A version of the approach is proposed for an analogous continuous-time process. The accuracy of the approximation is estimated.  The approach can be used in traffic models and models of statistical physics.
\end{abstract}
\hskip  15pt

\section*{1. Introduction} 

\hskip  16pt The concept of random exclusion process was introduced in [1]. A continuous-time totally asymmetric simple exclusion process
(TASEP) is defined as follows. Particles move on a one-dimensional lattice. There can be no more than one particle in a site simultaneously. The duration of intervals between attempts of a particle to move is distributed exponentially. An attempt is realized if the site ahead is vacant. In the case of discrete-time TASEP [2], at each step, with a prescribed probability, a particle moves onto a site in the direction of movement under the assumption that the site ahead is vacant. Exclusion processes are used [1]--[7] in models of statistical physics or traffic models 

In [3], an algorithm was proposed for the computation of steady probabilities for the continuous-time TASEP. In [4], an analogous algorithm was proposed for the computation of steady probabilities for discrete-time TASEP.

In [7], a continuous-time inhomogeneous asymmetric exclusion process is studied. The number of particles is finite. Particles can move in both directions. The rates of movement depend on particle. 

In [8], a discrete-time inhomogeneous TASEP on a lattice with open boundaries is considered. There are different types of particles. The probability of particle  movement depends on the particle type. An approach to approximate computation of the process characteristics is proposed. The accuracy of approximation for the case of two or
three particles is estimated.

In this paper, we estimate the accuracy of approximation for the approach proposed in [8] for discrete-time TASEP and propose an analogous approach for continuous-time TASEP. 

In [9], a discrete-time inhomogeneous TASEP on a closed lattice and a discrete-time inhomogeneous TASEP on an infinite lattice with finite number of particles are considered.

In Section 2, we describe  the process in consideration. In Section~3, we provide results regarding the ergodicity of the system. In Section~4, we describe an auxiliary homogeneous exclusion process. In Section~5, an approximate approach to evaluate the characteristics of the inhomogeneous process. In Section~6, we provide results of the approximate approach accuracy estimation. 
In Section~7, we propose a version of approach for a analogous continuous-time TASEP

\section*{2. Description of inhomogeneous process}

\hskip 16pt Suppose there is a lattice containing $N$ sites with indices $1,\dots,N.$ If, at time $t=0,1,2,\dots,$ the site~1 is vacant, then, with probability $\alpha,$ a particle arrives, and, in time $t+1,$ this particle will be in the site~1, $t=0,1,2,\dots$ Any particle belongs to one of $K$ types. With probability $a_k,$ a particle belongs to the type $k,$ $k=1,\dots,K,$ $a_1+\dots+a_k=1.$ If, at time $t,$ a particle of the type $k$ is in the site $i,$ then, with probability~$p_k,$ at time $t+1,$ the particle will occupy the site~$i+1$ under the assumption that, in
time $t,$ the site~$i+1$ is vacant,  $k=1,\dots,K,$ 
$i=0,1,\dots,N-1.$ If, at time~$t,$ a particle of the type~$k$ is in the site~$N,$ then,  with the probability~$\beta_k,$ the particle departs the system, and at time~$t+1,$ there is no particle in the site $T$,  $k=1,\dots,K,$ $t=0,1,2,\dots$

\section*{3. Ergodicity of the system}

\hskip 16pt Suppose $x=(x_1,\dots,x_N)$ is the state of the process such that $x_i=0$ if the site $i$ is vacant, and $x_i=k$ if the site $i$ is occupied by a particle of the type~$k,$ $i=1,\dots,N,$ $k=1,\dots,K.$ Let $x(t)$ be the state of the system at time $t\ge 0.$ The stochastic process $x(t)$ is a Markov chain with discrete time. In [8], it is proved that the process $x(t)$ is ergodic, i.e., there are steady state probabilities, and these probabilities do not depend on the initial state.   

Let the system be called the system $S.$ 

\section*{4. Description of homogeneous process}

\hskip 16pt Let us introduce the system $S^*.$ This system differs from the system $S$ by that the type of particles is unique. If the site ahead a particle is vacant, then the particle moves onto one site forward with probability $p^*$ such that 
$$p^*=\frac{1}{\sum\limits_{k=1}^K\frac{a_k}{p_k}},\eqno(1)$$
i.e. $p^*$ is the harmonic average of $p_1,\dots,p_K.$ 

Similarly, suppose that $\beta^*$ is the harmonic average of $\beta_1,\dots,\beta_K$,
$$\beta^*=\frac{1}{\sum\limits_{k=1}^K\frac{a_k}{\beta_k}}.\eqno(2)$$

Suppose $\eta(k)$ is a function defined on the set of numbers $0,1,\dots,K,$
$$\eta(k)=\left\{
\begin{array}{l}
0,\ k=0,\\
1,\ k>0.\\
\end{array}
\right.
$$
$(x_1^*,\dots,x_N^*)$ is the state of the system $S^*$ such that $x_i^*=0$ if there no particle in the site $i,$ and $x_i'=1$ if there is a particle in the  site $i,$ is $i=1,\dots,N.$

Let the state $(x_1^*,\dots,x_N^*)$ of the system $S^*$ correspond to the set of the system $S$ states $(x_1,\dots,x_N)$ such that $(\eta(x_1),\dots,\eta(x_N))=(x_1^*,\dots,x_N^*).$ Let the set be called the set $G(x_1^*,\dots,x_N^*).$

\section*{5. Approximate approach}

\hskip 16pt  In [8], an approach was proposed to evaluate the characteristics of the system $S.$ Denote by $\rho_i$ and $\rho_i^*$
the steady probability of the site $i$ occupancy (the particle density in site $i)$ for the system $S$ and $S^*$ respectively, $i=1,\dots,N.$ Denote by $J$ and $J^*$ the particle flow rate (average number of particles arrivals per a time unit) for the systems $S$ and $S^*$ respectively. 
 
 {\it The approximate values of the densities $\rho_1,\dots,\rho_N$ for the sites of the system $S$ and the rate $J$ for the system $S$ are assumed to be equal to the values of the densities $\rho_1^*,\dots,\rho_N^*$ respectively, and the rate $J$ is assumed to be equal to   $J^*.$ Besides, we consider the following characteristics:
$$\overline{\rho}=\frac{1}{N}\sum\limits_{i=1}^N\rho_i,$$
$\overline{\rho}$ is the average particle density,
$$\overline{\rho^*}=\frac{1}{N}\sum\limits_{i=1}^N\rho_i*^,$$
$$L=N\overline{\rho},$$
$L$ is the average number of particles in the system,
$$L^*=N\overline{\rho^*},$$ 
$$T=\frac{L}{J}$$ 
(Little's formula), $T$ is the average sojourn time for a particle in the system,
$$T^*=\frac{L^*}{J^*},$$
$$\overline{v}=\frac{N}{T}=\frac{J}{\overline{\rho}},$$
$\overline{v}$ is the average velocity of particles,
$$\overline{v^*}=\frac{N^*}{T^*}=\frac{J^*}{\overline{\rho^*}},$$
$\overline{\rho^*},$ $L^*,$ $T^*,$ $\overline{v^*}$ are approximate values of related characteristics.
}

In [8], it has been proved that this approach is exact in the case $n=2,$
 $\beta_1=\dots=\beta_{K}.$

\section*{6. Estimation of approximation accuracy}

\hskip 16pt In [8], the approximation accuracy was estimated for the cases of $N=2$ and $N=3.$

The results of the approximation accuracy evaluation for the case of two sites are provided in Table~1. Exact values (up to rounding) were found from the system
of equations for stationary state probabilities. Approximate values were computed according to the proposed approach. The upper parts of cells contain exact values. The lower parts of cells contain approximate values. 
\vskip 10pt
{\normalsize
\vskip 20pt
{\bf Table 1 [8].} Exact and approximate values of synchronous process characteristics, $N=2,$ $K=2.$
\vskip 10pt
\begin{tabular}{|c|c|c|c|c|c|c|c|c|c|c|c|c|c|c|c|}
\hline
&$\alpha$&$a_1$&$a_2$&$p_1$&$p_2$&$\beta_1$&$\beta_2$&
$\rho_1$&$\rho_2$&$J$&$\overline{\rho}$&$L$&$T$&$v$\\
\hline
\hline
1&$\frac{2}{5}$&$\frac{3}{7}$&$\frac{4}{7}$&$\frac{3}{5}$&$\frac{4}{5}$&$\frac{3}{10}$&$\frac{2}{5}$&$\frac{0.5149}{0.5142}$&$\frac{0.5544}{0.5552}$&
$\frac{0.1940}{0.1943}$&$\frac{0.5347}{0.5347}$&$\frac{1.0693}{1.0694}$&$\frac{5.512}{5.504}$&$\frac{0.3628}{0.3637}$\\
\hline
\hline
2&$\frac{1}{5}$&$\frac{2}{5}$&$\frac{3}{5}$&$\frac{2}{5}$&$\frac{3}{5}$&$\frac{1}{5}$&$\frac{3}{10}$&$\frac{0.4135}{0.4118}$&$\frac{0.4692}{0.4706}$&$\frac{0.1173}{0.1176}$&$\frac{0.4414}{0.4412}$&$\frac{0.8827}{0.8824}$&$\frac{7.525}{7.503}$&
$\frac{0.2658}{0.2666}$\\
\hline
\hline
3&$\frac{1}{5}$&$\frac{1}{3}$&$\frac{2}{3}$&$\frac{2}{5}$&$\frac{4}{5}$&$\frac{1}{5}$&$\frac{2}{5}$&$\frac{0.3583}{0.3529}$&$\frac{0.4278}{0.4314}$&$\frac{0.1283}{0.1294}$&$\frac{0.3931}{0.3922}$&
$\frac{0.7861}{0.7843}$&$\frac{6.127}{6.038}$&$\frac{0.3264}{0.3312}$&\\
\hline
\hline
4&$\frac{8}{25}$&$\frac{3}{4}$&$\frac{1}{4}$&$\frac{12}{25}$&$\frac{18}{25}$&$\frac{9}{25}$&$\frac{11}{25}$&$\frac{0.4752}{0.4749}$&$\frac{0.4453}{0.4455}$&$\frac{0.1679}{0.1680}$&$\frac{0.4603}{0.4602}$&
$\frac{0.9205}{0.9204}$&
$\frac{5.482}{5.479}$&
$\frac{0.3648}{0.3650}$&
 
\\
\hline
\hline
5&$\frac{8}{25}$&$\frac{3}{4}$&$\frac{1}{4}$&$\frac{12}{25}$&$\frac{18}{25}$&$\frac{1}{5}$&$\frac{2}{5}$&$\frac{0.5744}{0.5723}$&$\frac{0.5958}{0.5988}$&$\frac{0.1362}{0.1369}$&$\frac{0.5851}{0.5856}$&$\frac{1.1702}{1.1711}$&$\frac{8.592}{8.554}$&$\frac{0.2328}{0.2358}$\\
\hline
\end{tabular}
}
\vskip 10pt
 In [8],  the following results of approximation accuracy evaluation for the case of three sites were provided.
 \vskip 5pt 
Suppose $N=3,$ $K=2,$ $\alpha=\frac{1}{5},$ $a_1=\frac{2}{5},$ $a_2=\frac{3}{5},$  $p_1=\frac{2}{5},$ $p_2=\frac{3}{5},$ 
 $\beta_1=\frac{1}{5},$ $\beta_2=\frac{3}{10}.$
The exact values are
$$\rho_1=0.3988,\ \rho_2=0.4374,\ \rho_3=0.4809,\  J=0.1202,$$
$$\overline{\rho}=0.4390,\ L=1.3171,\  T=10.958,\ 
\overline{v}=0.2738.$$

The approximate values are
$$\rho_1^*=0.3952,\ \rho_2^*=0.4355,\ \rho_3^*=0.4838,\ J^*=0.1210,$$
$$\overline{\rho^*}=0.4382 ,\ L^*=1.3145,\ T^*=10.864,\ \overline{v^*}=0.2761.$$

Let us to provide results of the approximation accuracy evaluation for the cases $N=10$ and $N=100.$ Suppose $\alpha=\frac{1}{10},$ $a_1=\frac{1}{3},$ $a_2=\frac{2}{3},$  $p_1=\frac{2}{15},$ $p_2=\frac{4}{15}.$

Suppose $N=10,$ $K=2,$ 
 $\beta_1=\frac{1}{15},$ $\beta_2=\frac{2}{15}.$ Results of simulation are shown in Table 2. Simulation interval is 1000000 time units.
\vskip 10pt
{\bf Table 2.} Exact and approximate values of density, 
$N=10,$ $K=2.$
\vskip 10pt
\begin{tabular}{|c|c|c|c|c|c|c|c|c|c|c|}
\hline
\hline
Site index&$\rho$&$\rho^*$\\
\hline
\hline
1&$2$&$3$\\
\hline
\hline
1&$\rho=0.5091$&$\rho^*=0.4830$\\
\hline
\hline
2&$\rho=0.5240$&$\rho^*=0.4864$\\
\hline
\hline
3&$\rho=0.5329$&$\rho^*=0.4898$\\
\hline
\hline
4&$\rho=0.5369$&$\rho^*=0.4930$\\
\hline
\hline
5&$\rho=0.5376$&$\rho^*=0.5017$\\
\hline
\hline
6&$\rho=0.5350$&$\rho^*=0.5017$\\
\hline
\hline
7&$\rho=0.5356$&$\rho^*=0.5024$\\
\hline
\hline
8&$\rho=0.5198$&$\rho^*=0.5053$\\
\hline
\hline
9&$\rho=0.5080$&$\rho^*=0.5145$\\
\hline
\hline
10&$\rho=0.4949$&$\rho^*=0.5196$\\
\hline
\hline
\end{tabular}
\vskip 10pt
The values of the rate are $J=0.0491,$ $J^*=0.0517.$ 

The average density is
$$\overline{\rho}=\frac{1}{N}\sum\limits_{i=1}^N\rho_i=0.5217.$$
The approximate average of densities is
$$\overline{\rho^*}=\frac{1}{N}\sum\limits_{i=1}^N\rho_i^*=0.5006.$$

The average number $L$ of particles in the system is $L=N\overline{\rho}=5.217.$ The approximate average number $L^*$ of particles in the system is $L^*=N\overline{\rho}=5.006.$

According to Little's formula for queueing systems, the average number of particles in the system is equal to the product of the rate $J$ and the average sojourn time $T$ for a particle in the system $L=JT.$ Therefore,
$$T=\frac{L}{J}=106.25.$$  
The approximate value $T^*$ of sojourn time is 
$$T^*=\frac{L^*}{J^*}=96.825.$$ 

The average velocity $\overline v$ on the lattice is
$$\overline v=\frac{N}{T}=\frac{J}{\overline{\rho}}=0.0941.$$

The  approximate value $\overline v^*$ of the average velocity is
$$\overline v^*=\frac{N}{T^*}=\frac{J^*}{\overline{\rho^*}}=0.1033.$$

\vskip 20pt
Suppose $N=100,$ $K=2,$ $\alpha=\frac{1}{10},$ $a_1=\frac{1}{3},$ $a_2=\frac{2}{3},$  $p_1=\frac{2}{15},$ $p_2=\frac{4}{15},$ 
 $\beta_1=\frac{1}{15},$ $\beta_2=\frac{2}{15}.$ Results of the simulation are shown in Table 3. Simulation interval is 1000000 time units.
\vskip 10pt
{\bf Table 3.} Values of density, 
$N=10,$ $K=2.$
\vskip 10pt
\begin{tabular}{|c|c|c|c|c|c|c|c|c|c|c|}
\hline
\hline
Site index&$\rho$&$\rho^*$\\
\hline
\hline
1&$\rho=0.5297$&$\rho^*=0.4767$\\
\hline
\hline
2&$\rho=0.5586$&$\rho^*=0.4817$\\
\hline
\hline
3&$\rho=0.5722$&$\rho^*=0.4831$\\
\hline
\hline
4&$\rho=0.5824$&$\rho^*=0.4809$\\
\hline
\hline
5&$\rho=0.5848$&$\rho^*=0.4809$\\
\hline
\hline
6&$\rho=0.5867$&$\rho^*=0.4838$\\
\hline
\hline
7&$\rho=0.5882$&$\rho^*=0.4847$\\
\hline
\hline
8&$\rho=0.5881$&$\rho^*=0.4841$\\
\hline
\hline
9&$\rho=0.5888$&$\rho^*=0.4834$\\
\hline
\hline
10&$\rho=0.5878$&$\rho^*=0.4842$\\
\hline
\hline
11&$\rho=0.5879$&$\rho^*=0.4802$\\
\hline
\hline
12&$\rho=0.5889$&$\rho^*=0.4833$\\
\hline
\hline
13&$\rho=0.5865$&$\rho^*=0.4829$\\
\hline
\hline
14&$\rho=0.5859$&$\rho^*=0.4829$\\
\hline
\hline
15&$\rho=0.5869$&$\rho^*=0.4836$\\
\hline
\hline
16&$\rho=0.5861$&$\rho^*=0.4834$\\
\hline
\hline
17&$\rho=0.5842$&$\rho^*=0.4842$\\
\hline
\hline
18&$\rho=0.5840$&$\rho^*=0.4833$\\
\hline
\hline
\end{tabular}

\begin{tabular}{|c|c|c|c|c|c|c|c|c|c|c|}
\hline
\hline
19&$\rho=0.5823$&$\rho^*=0.4850$\\
\hline
\hline
20&$\rho=0.5823$&$\rho^*=0.4828$\\
\hline
\hline
21&$\rho=0.5807$&$\rho^*=0.4838$\\
\hline
\hline
22&$\rho=0.5787$&$\rho^*=0.4842$\\
\hline
\hline
23&$\rho=0.5757$&$\rho^*=0.4852$\\
\hline
\hline
24&$\rho=0.5756$&$\rho^*=0.4842$\\
\hline
\hline
25&$\rho=0.5768$&$\rho^*=0.4857$\\
\hline
\hline
26&$\rho=0.5776$&$\rho^*=0.4882$\\
\hline
\hline
27&$\rho=0.5771$&$\rho^*=0.4865$\\
\hline
\hline
28&$\rho=0.5771$&$\rho^*=0.4857$\\
\hline
\hline
29&$\rho=0.5766$&$\rho^*=0.4862$\\
\hline
\hline
30&$\rho=0.5748$&$\rho^*=0.4857$\\
\hline
\hline
31&$\rho=0.5748$&$\rho^*=0.4878$\\
\hline
\hline
32&$\rho=0.5736$&$\rho^*=0.4890$\\
\hline
\hline
33&$\rho=0.5721$&$\rho^*=0.4886$\\
\hline
\hline
34&$\rho=0.5706$&$\rho^*=0.4897$\\
\hline
\hline
33&$\rho=0.5721$&$\rho^*=0.4886$\\
\hline
\hline
34&$\rho=0.5706$&$\rho^*=0.4897$\\
\hline
\hline
35&$\rho=0.5696$&$\rho^*=0.4904$\\
\hline
\hline
36&$\rho=0.5686$&$\rho^*=0.4914$\\
\hline
\hline
37&$\rho=0.5659$&$\rho^*=0.4910$\\
\hline
\hline
38&$\rho=0.5632$&$\rho^*=0.4925$\\
\hline
\hline
39&$\rho=0.5632$&$\rho^*=0.4903$\\
\hline
\hline
40&$\rho=0.5641$&$\rho^*=0.4918$\\
\hline
\hline
41&$\rho=0.5635$&$\rho^*=0.4879$\\
\hline
\hline
42&$\rho=0.5638$&$\rho^*=0.4894$\\
\hline
\hline
43&$\rho=0.5654$&$\rho^*=0.4932$\\
\hline
\hline
44&$\rho=0.5628$&$\rho^*=0.4912$\\
\hline
\hline
45&$\rho=0.5614$&$\rho^*=0.4948$\\
\hline
\hline
46&$\rho=0.5615$&$\rho^*=0.4964$\\
\hline
\hline
47&$\rho=0.5630$&$\rho^*=0.4964$\\
\hline
\hline
48&$\rho=0.5630$&$\rho^*=0.4999$\\
\hline
\hline
49&$\rho=0.5596$&$\rho^*=0.4998$\\
\hline
\hline
50&$\rho=0.5593$&$\rho^*=0.4983$\\
\hline
\hline
51&$\rho=0.5589$&$\rho^*=0.4986$\\
\hline
\hline
52&$\rho=0.5579$&$\rho^*=0.4996$\\
\hline
\hline
53&$\rho=0.5545$&$\rho^*=0.4999$\\
\hline
\hline
54&$\rho=0.5567$&$\rho^*=0.4980$\\
\hline
\hline
55&$\rho=0.5542$&$\rho^*=0.4964$\\
\hline
\hline
\end{tabular}

\newpage

\begin{tabular}{|c|c|c|c|c|c|c|c|c|c|c|}
\hline
\hline
56&$\rho=0.5548$&$\rho^*=0.4975$\\
\hline
\hline
57&$\rho=0.5537$&$\rho^*=0.4977$\\
\hline
\hline
58&$\rho=0.5534$&$\rho^*=0.4995$\\
\hline
\hline
59&$\rho=0.5545$&$\rho^*=0.5004$\\
\hline
\hline
60&$\rho=0.5525$&$\rho^*=0.4977$\\
\hline
\hline
61&$\rho=0.5525$&$\rho^*=0.4980$\\
\hline
\hline
62&$\rho=0.5529$&$\rho^*=0.4982$\\
\hline
\hline
63&$\rho=0.5531$&$\rho^*=0.4980$\\
\hline
\hline
64&$\rho=0.5508$&$\rho^*=0.5004$\\
\hline
\hline
65&$\rho=0.5511$&$\rho^*=0.5015$\\
\hline
\hline
66&$\rho=0.5527$&$\rho^*=0.5006$\\
\hline
\hline
67&$\rho=0.5511$&$\rho^*=0.5020$\\
\hline
\hline
68&$\rho=0.5524$&$\rho^*=0.5001$\\
\hline
\hline
69&$\rho=0.5494$&$\rho^*=0.5024$\\
\hline
\hline
70&$\rho=0.5481$&$\rho^*=0.5055$\\
\hline
\hline
71&$\rho=0.5480$&$\rho^*=0.5050$\\
\hline
\hline
72&$\rho=0.5473$&$\rho^*=0.5072$\\
\hline
\hline
73&$\rho=0.5467$&$\rho^*=0.5087$\\
\hline
\hline
74&$\rho=0.5450$&$\rho^*=0.5066$\\
\hline
\hline
75&$\rho=0.5450$&$\rho^*=0.5078$\\
\hline
\hline
76&$\rho=0.5469$&$\rho^*=0.5088$\\
\hline
\hline
77&$\rho=0.5445$&$\rho^*=0.5114$\\
\hline
\hline
78&$\rho=0.5433$&$\rho^*=0.5134$\\
\hline
\hline
79&$\rho=0.5386$&$\rho^*=0.5127$\\
\hline
\hline
80&$\rho=0.5377$&$\rho^*=0.5114$\\
\hline
\hline
81&$\rho=0.5351$&$\rho^*=0.5097$\\
\hline
\hline
82&$\rho=0.5369$&$\rho^*=0.5098$\\
\hline
\hline
83&$\rho=0.5344$&$\rho^*=0.5096$\\
\hline
\hline
84&$\rho=0.5354$&$\rho^*=0.5097$\\
\hline
\hline
85&$\rho=0.5315$&$\rho^*=0.5137$\\
\hline
\hline
86&$\rho=0.5296$&$\rho^*=0.5133$\\
\hline
\hline
87&$\rho=0.5292$&$\rho^*=0.5150$\\
\hline
\hline
88&$\rho=0.5276$&$\rho^*=0.5170$\\
\hline
\hline
89&$\rho=0.5233$&$\rho^*=0.5168$\\
\hline
\hline
90&$\rho=0.5233$&$\rho^*=0.5168$\\
\hline
\hline
91&$\rho=0.5210$&$\rho^*=0.5156$\\
\hline
\hline
92&$\rho=0.5195$&$\rho^*=0.5167$\\
\hline
\hline
93&$\rho=0.5156$&$\rho^*=0.5180$\\
\hline
\hline
94&$\rho=0.5157$&$\rho^*=0.5190$\\
\hline
\hline
95&$\rho=0.5136$&$\rho^*=0.5187$\\
\hline
\hline
\end{tabular}

\newpage
\begin{tabular}{|c|c|c|c|c|c|c|c|c|c|c|}
\hline
\hline
96&$\rho=0.5208$&$\rho^*=0.5208$\\
\hline
\hline
97&$\rho=0.5048$&$\rho^*=0.5217$\\
\hline
\hline
98&$\rho=0.4968$&$\rho^*=0.5224$\\
\hline
\hline
99&$\rho=0.4909$&$\rho^*=0.5268$\\
\hline
\hline
100&$\rho=0.4756$&$\rho^*=0.5280$\\
\hline
\hline
\end{tabular}
\vskip 10pt
The values of the rate are $J=0.0941,$ $J^*=0.1047.$

The average density is
$$\overline{\rho}=0.5581.$$
The approximate average  value of density
$$\overline{\rho^*}=0.5029.$$

The average number $L$ of particles in the system is $L=55.81.$
The average number $L^*$ of particles in the system is $L^*=50.29.$

The average sojourn time $T$ is 
$$T=593.4.$$  
The approximate value $T^*$ of sojourn time is 
$$T^*=480.5.$$ 

The average velocity $\overline v$ on the lattice is
$$\overline v=0.1685.$$
The approximate value  $\overline v$ of average velocity  is
$$\overline{v^*}=0.2081.$$

\vskip 20pt
In the case of three sites, the following results were obtained. 
 \vskip 5pt 
Suppose $N=3,$ $\alpha=\frac{1}{5},$ $a_1=\frac{2}{5},$ $a_2=\frac{3}{5},$ $\mu_1=\frac{2}{5},$ $\mu_2=\frac{3}{5},$ 
 $\beta_1=\frac{1}{5},$ $\beta_2=\frac{3}{10}.$
Exact values are
$$\rho_1=0.4262,\  \rho_2=0.4419,\ \rho_3=0.4591,\  J=0.1148,$$
$$\overline{\rho}=0.4424,\  L=1.327,\ T=11.56,\  \overline{v}=0.2595.$$
Approximate values are
$$\rho_1^*=0.4233,\  \rho_2^*=0.4409,\  \rho_3^*=0.4613,\  J^*=0.1153,$$
$$\overline{\rho^*}=0.4418,\  L^*=1.326,\  T^*=11.50,\  \overline{v^*}=0.2609.$$
\vskip 5pt
Suppose $N=3,$ $\alpha=\frac{1}{5},$ $a_1=\frac{1}{3},$ $a_2=\frac{2}{3},$
$\mu_1=\frac{1}{3},$ $\mu_2=\frac{2}{3},$ 
 $\beta_1=\frac{2}{5},$ $\beta_2=\frac{4}{5}.$
Exact values are:
$$\rho_1=0.3709,\ \rho_2=0.3948,\  \rho_3=0.4193,\  J=0.1258,$$
$$\overline{\rho}=0.3950,\  L=1.185,\ T=9.412,\ \overline{v}=0.3187.$$
Approximate values:
$$\rho_1^*=0.3628,\  \rho_2^*=0.3894,\  \rho_3^*=0.4248,\ J^*=0.1274,$$
$$\overline{\rho^*}=0.3923,\  L^*=1.177,\  T^*=9.239,\ \overline{v^*}=0.3247.$$

\section*{7. Continuous-time inhomogeneous process}

\hskip 16pt In this section, we propose a version of the approximate approach
for a continuous-time inhomogeneous TASEP. 

This process differs from the discrete-time process by that the time scale is continuous, and the time-intervals between attempts of particle $i$ to move are distributed exponentially with rate $\mu_i,$ $i=1,\dots,N.$ 

The proof of the continuous-time process ergodicity is the same as for the discrete-time process.

The results of the approximation accuracy evaluation for the case of two sites are provided
in Table~1. Exact values were found from the system
of equations for stationary state probabilities. Approximate values were computed according to the proposed approach. The upper parts of cells contain exact values. The lower parts of cells contain approximate values. 

\vskip 10pt
{\bf Table 4.} Exact and approximate values of asynchronous process characteristics
$N=2.$
\vskip 10pt

{\small
\begin{tabular}{|c|c|c|c|c|c|c|c|c|c|c|c|c|c|c|}
\hline
No.& $\alpha$&$a_1$&$a_2$&$\mu_1$&$\mu_2$&$\beta_1$&$\beta_2$&
$\rho_1$&$\rho_2$&$J$&$\overline{\rho}$&$L$&$T$&$\overline{v}$\\
\hline
1&$\frac{2}{5}$&$\frac{3}{7}$&$\frac{4}{7}$&$\frac{3}{5}$&$\frac{4}{5}$&$\frac{3}{10}$&$\frac{2}{5}$&
$\frac{0.5421}{0.5415}$&$\frac{0.5234}{0.5240}$&
$\frac{0.1832}{0.1834}$&$\frac{0.5328}{0.5328}$&$\frac{1.066}{1.066}$&$\frac{5.816}{5.810}$
&$\frac{0.3439}{0.3442}$\\
\hline
2&$\frac{1}{5}$&$\frac{2}{5}$&$\frac{3}{5}$&$\frac{2}{5}$&$\frac{3}{5}$&$\frac{1}{5}$&$\frac{3}{10}$&
$\frac{0.4318}{0.4304}$&$\frac{0.4545}{0.4557}$&
$\frac{0.1136}{0.1139}$&$\frac{0.4432}{0.4431}$&$\frac{0.8863}{0.8861}$
&$\frac{7.802}{7.780}$&$\frac{0.2563}{0.2571}$ 
 \\
\hline
3&$\frac{1}{5}$&$\frac{1}{3}$&$\frac{2}{3}$&$\frac{2}{5}$&$\frac{4}{5}$&$\frac{1}{5}$&$\frac{2}{5}$&$\frac{0.3793}{0.3750}$&
$\frac{0.4138}{0.4167}$&$\frac{0.1241}{0.1250}$&$\frac{0.3966}{0.3959}$
&$\frac{0.7931}{0.7917}$&$\frac{6.391}{6.334}$&$\frac{0.3129}{0.3158}$\\
\hline
4&$\frac{8}{25}$&$\frac{3}{4}$&$\frac{1}{4}$&$\frac{12}{25}$&$\frac{18}{25}$&$\frac{9}{25}$&$\frac{11}{25}$&$\frac{0.5003}{0.5001}$&$
\frac{0.4239}{0.4241}$&$\frac{0.1599}{0.1600}$&$\frac{0.4621}{0.4621}$&$\frac{0.9242}{0.9242}$&$
\frac{5.779}{5.776}$&$\frac{0.3461}{0.3463}$\\
\hline
5&$\frac{8}{25}$&$\frac{3}{4}$&$\frac{1}{4}$&$\frac{12}{25}$&$\frac{18}{25}$&$\frac{1}{5}$&$\frac{2}{5}$&$\frac{0.5899}{0.5881}$
&$\frac{0.5741}{0.5767}$&$\frac{0.1312}{0.1318}$
&$\frac{0.5820}{0.5824}$&$\frac{1.164}{1.165}$
&$\frac{8.872}{8.838}$&$\frac{0.2254}{0.2263}$
\\
\hline
\end{tabular}
}
\vskip 20pt
For the case of three sites, the proposed approximate approach provides the following results.  

\vskip 20pt
{\bf Table 5.} Exact and approximate values of asynchronous process characteristics $N=3.$
\vskip 10pt

{\small
\begin{tabular}{|c|c|c|c|c|c|c|c|c|c|c|c|c|c|c|c|}
\hline
No. &$\alpha$&$a_1$&$a_2$&$\mu_1$&$\mu_2$&$\beta_1$&$\beta_2$&
$\rho_1$&$\rho_2$&$\rho_3$&$J$&$\overline{\rho}$&$L$&$T$&$\overline{v}$\\
\hline
1&$\frac{2}{5}$&$\frac{3}{7}$&$\frac{4}{7}$&$\frac{3}{5}$&$\frac{4}{5}$&$\frac{3}{10}$&$\frac{2}{5}$&
$\frac{0.5467}{0.5455}$&$\frac{0.5324}{0.5319}$&$\frac{0.5181}{0.5194}$&
$\frac{0.1813}{0.1818}$&$\frac{0.5324}{0.5323}$
&$\frac{1.597}{1.597}$&$\frac{8.810}{8.783}$&$\frac{0.3405}{0.3416}$
\\
\hline
2&$\frac{1}{5}$&$\frac{2}{5}$&$\frac{3}{5}$&$\frac{2}{5}$&$\frac{3}{5}$&$\frac{1}{5}$&$\frac{3}{10}$&
$\frac{0.4267}{0.4234}$&$\frac{0.4425}{0.4409}$&
$\frac{0.4591}{0.4613}$&
$\frac{0.1148}{0.1153}$&
$\frac{0.4428}{0.4419}$&$\frac{1.328}{1.326}$&
$\frac{11.51}{11.50}$&
$\frac{0.1148}{0.1153}$
\\
\hline
3&$\frac{1}{5}$&$\frac{1}{3}$&$\frac{2}{3}$&$\frac{2}{5}$&$\frac{4}{5}$&$\frac{1}{5}$&$\frac{2}{5}$&$\frac{0.3710}{0.3628}$&$\frac{0.3948}{0.3894}$&$\frac{0.4193}{0.4248}$&$\frac{0.1258}{0.1274}$&
$\frac{0.5816}{0.5808}$&$\frac{1.745}{1.743}$&
$\frac{9.349}{9.239}$&$\frac{0.3209}{0.3247}$

\\
\hline
4&$\frac{8}{25}$&$\frac{3}{4}$&$\frac{1}{4}$&$\frac{12}{25}$&$\frac{18}{25}$&$\frac{9}{25}$&$\frac{11}{25}$&$\frac{0.5207}{0.5197}$&$
\frac{0.4673}{0.4664}$&$\frac{0.4071}{0.4075}$& $\frac{0.1534}{0.1537}$
&$\frac{0.4650}{0.4645}$&$\frac{1.395}{1.394}$&$\frac{9.095}{9.067}$
&$\frac{0.3299}{0.3309}$
\\
\hline
5&$\frac{8}{25}$&$\frac{3}{4}$&$\frac{1}{4}$&$\frac{12}{25}$&$\frac{18}{25}$&$\frac{1}{5}$&$\frac{2}{5}$&$\frac{0.5934}{0.5881}$&$
\frac{0.5819}{0.5810}$&$\frac{0.5694}{0.5734}$&$\frac{0.1311}{0.1318}$
&$\frac{0.5816}{0.5808}$&$\frac{1.745}{1.743}$&$\frac{13.31}{13.22}$
&$\frac{0.2254}{0.2269}$\\
\hline
\end{tabular}
}

\section*{8. Conclusion}

\hskip 16pt We consider an inhomogeneous exclusion process on an open one-dimensional is considered. In [8], for the  discrete-time version of the process an approximate approach to compute characteristics of the process is proposed, and the accuracy of the approach is estimated for the cases of 2 and 3 sites. In this paper, the accuracy of the approach is estimated for the cases of 10 and 100 sites. We also propose an analogous approximate approach to compute characteristics of the continuous-time version of the process.

\section*{References}

\hskip 18pt 1. Spitzer F. Interaction of Markov processes. In Random Walks, Browing Motion, and Interacting Particle Systems: A Festschrift in Honor of Frank Spitzer. -- Boston, MA : Birkhäuser Boston, 1991, pp. 66--110.

2. Evans M.R., Rajewsky N., Speer E.R. Exact solution of a cellular automaton for traffic. Journal of Statistical Physics,1999, vol. 95, pp.~45--96. DOI: 10.1023/A:1004521326456

3. Derrida B. An exactly soluble non-equilibrium system: The asymmetric simple exclusion process. Physics Reports, 1998, vol.~301, no.~1--3, pp.~65--83. DOI: 10.1016/S0370-1573(98)00006-4 .

4. Schadschneider A., Schreckenberg M. Cellular automation models and traffic flow. Journal of Physics A: Mathematical and General, 1993, vol.~26, no.~15, pp.~L679. DOI: 10.1088/0805-0305-4470-26/15/011

5. Kanai~M., Nishinari~K., Tokihiro~T. (2006). Exact solution and asymptotic behaviour of the asymmetric simple exclusion process on a ring. Journal of Physics A: Mathematical and General, 39(29), 9071.
DOI: 10.1088/0305-4470/39/29/004

6. Blank M.L. Metric properties of discrete time exclusion type processes in continuum. Journal of Statistical Physics, 2010, vol.~140, no.~1, pp.~170--197. DOI: 10.1007/s10955-010-9983-y

7. Malyshev V., Menshikov M., Popov S., Wade A. Dynamics of finite inhomogeneous particle systems with exclusion interaction. Journal of Statistical Physics {\bf 190}(11), 184 (2023). https://doi.org/10.1007/s10955-023-03190-8

8. Yashina M.V., Tatashev A.G. (2024). Synchronous Heterogeneous Exclusion Processes on Open Lattice. arXiv preprint arXiv:2411.12419.

\end{document}